\documentstyle[aps,prl,epsf]{revtex}
\tighten

\begin{document}
\newcommand{\gsim}{\gtrsim}
\newcommand{\lsim}{\lesssim}
\newcommand{\psim}{\mbox{\raisebox{-1.0ex}{$~\stackrel{\textstyle \propto}
{\textstyle \sim}~$ }}}
\newcommand{\vect}[1]{\mbox{\boldmath${#1}$}}
\newcommand{\lmk}{\left(}
\newcommand{\rmk}{\right)}
\newcommand{\lnk}{\left\{ }
\newcommand{\nn}{\nonumber}
\newcommand{\rnk}{\right\} }
\newcommand{\lkk}{\left[}
\newcommand{\rkk}{\right]}
\newcommand{\lla}{\left\langle}
\newcommand{\p}{\partial}
\newcommand{\rra}{\right\rangle}
\newcommand{\beq}{\begin{equation}}
\newcommand{\eeq}{\end{equation}}
\newcommand{\beqa}{\begin{eqnarray}}
\newcommand{\eeqa}{\end{eqnarray}}
\newcommand{\lab}{\label}
\newcommand{\sol}{M_\odot}

\draft
\title{Proposal for Determining the  Total Masses of Eccentric 
 Binaries\\
 Using Signature of
 Periastron Advance  in  
Gravitational Waves}

\author{Naoki Seto
}
\address{Department of Earth and Space Science, Osaka
University, Toyonaka 560-0043, Japan}

\maketitle

\begin{abstract}
 We propose a new method for determining  total masses of low frequency
 eccentric binaries  (such as, neutron star binaries with orbital
 frequency  $f\gsim  10^{-3}$Hz) from  their gravitational waves.  
 In this  method we use the frequency shift caused by  periastron
 advance, and it works even at low frequency band where  chirp signal
 due to radiation reaction  is difficult to be measured. It is shown
 that the total masses of several Galactic neutron star binaries might
 be measured accurately (within a few percent error) by  LISA with
 operation period of  $\sim 10$ years.  
\end{abstract}

\section{Introduction}
The Galactic binaries such as close white dwarf binaries (CWDBs) or
neutron star binaries (NBs) are promising targets of the Laser
Interferometer Space Antenna (LISA, http://lisa.jpl.nasa.gov,
\cite{lisa}). One of the 
important aim of gravitational-wave astronomy is to extract out
information of sources that emit gravitational radiation. In the case of a
Galactic binary we want to know 
masses of two stars, orbital parameters (semi-major axis, eccentricity,
inclination), distance to the binary and so on. At the final in-spiral
phase ({\it e.g.} the last three minutes of NBs) we can, in principle,
 determine various
parameters by fitting time evolution of wave signal with post-Newtonian
expansion \cite{Cutler:1993tc}. But at LISA band many 
compact binaries evolve very slowly and situation is largely different.

It is observationally known that NBs (and some NS-WD binaries) with
orbital period $\lsim 1$day have large eccentricities \cite{brown}. This
is explained by the ``kick'' effect and instantaneous mass loss at birth
of a neutron star, in 
contrast to CWDBs whose orbits are circulaized by strong tidal interaction
during mass transfer phases.  As the eccentricity 
$e$ decreases with increase of the orbital 
frequency  due to radiation reaction \cite{peters},  its
effect 
is supposed to  be  
negligible for most high-frequency sources that will be
searched by ground-based
detectors (TAMA300, GEO600, LIGO and VIRGO). But  the eccentricity of
NBs can be $e\sim 0.1$ at frequency $f \sim  10^{-3}$Hz as expected from
PSR B1913+16 \cite{1913},  and it might cause observable effects for
LISA.  

Frequency series of gravitational wave from an eccentric binary is
affected by the  periastron advance 
\cite{moreno} (see Ref.\cite{1913} for radio observation of binary
pulsars). Using this 
fact we  
might obtain information of binaries even at low frequency band where
chirp signal would be difficult to be measured. In this Letter we propose
a new method for determining binary parameters (mainly total mass),  and
investigate its  
prospect for studying Galactic NBs with LISA.
 
\section{Gravitational Wave from Elliptical Orbit}
Let us study  gravitational waves from elliptical orbits following
Ref.\cite{moreno} (see also \cite{Wahlquist:1987rx}). 
With quadrupole formula of gravitational radiation 
\cite{gravitation} two polarization modes in
TT-gauge are written as follows 
\beqa
h_{\times}&=&-h_0 \cos\Theta \sum_n \lkk \frac{S_n-C_n}2  \sin
(2\pi f_n t+2\Phi)   
+ \frac{S_n+C_n}2  \sin (2\pi f_n t-2\Phi)  \rkk, \label{hc}\\
h_{+}&=&-\frac12 h_0  \sum_n \Bigg\{ \sin^2\Theta A_n  \cos
(2\pi f_nt)  +
(1+\cos^2\Theta )\times \nonumber\\ 
& & \lkk \frac{S_n-C_n}2  \cos (2\pi f_n t+2\Phi) 
+ \frac{S_n+C_n}2  \cos (2\pi f_n t-2\Phi)  \rkk \Bigg\},\label{hp}
\eeqa 
where $\Theta$ represents
direction of the orbital angular momentum
(inclination) and $\Phi$ represents  direction of the periastron in
the orbital plane (we put $\Phi=0$ when the periastron is on the plane
determined by two vectors: the orbital angular momentum vector and the
 direction
vector to the 
observer \cite{moreno}).   The amplitude $h_0$ is given as
$h_0=4(2\pi)^{-2/3}G^{5/3}c^{-4}m_{chirp}^{5/3}r^{-1} (P_b)^{-2/3}$
($m_{chirp}$: 
chirp mass, $r$: distance to  the binary, $P_b$: orbital period from
periastron to periastron).
We have defined the
 frequency series  $f_n\equiv P_b^{-1} n$. 
The angle $\Phi$ changes due to the 
 periastron advance and we can denote it
as $2\Phi=\delta f \times t$ by
adjusting origin of the  time coordinate
$t$. Thus the periastron advance causes frequency  shift $\delta f$ that
 is given in terms of the 
total mass $m_{total}$ and the  eccentricity $e$ 
 as $\delta f=6(2\pi)^{2/3}(P_b)^{-5/3}G^{2/3}c^{-2}
m_{total}^{2/3}(1-e^2)^{-1}$ (twice of  the frequency for the periastron
advance \cite{gravitation}).  The frequency $f_n$  now splits 
into a triplet $(f_n-\delta f, f_n, f_n+\delta f)$. Effect of the angle
$2\Phi$ is related to  rotation of a coordinate system and appears in the
same form for every $n$-mode. This is an essential point.  
 The simple replacement $2\Phi\to
2\delta f\times t$ into  expressions
(\ref{hc}) and (\ref{hp})  corresponds to  an
approximation that neglects terms of $O(P_b \delta f)\ll 1$\cite{moreno}.
  The   
coefficients $S_n$, $C_n$ and $A_n$ are given by the $n$-th Bessel function
$J_n(x)$
and its derivative $J^{'}_n (x)\equiv\p_x J_n (x) $ as 
\beqa
A_n&=&J_n(ne),~~~
S_n=-\frac{2(1-e^2)^{1/2}}{e} J^{'}_n(ne)+\frac{2n(1-e^2)^{3/2}}{e^2}
J_n (ne), \\
C_n&=&-\frac{2-e^2}{e^2} J_n(ne)+\frac{2(1-e^2)}{e}
J^{'}_n (ne) .
\eeqa
We can expand these coefficients around $e=0$ (circular orbit) and find
that only $n\le 3$ 
modes have  terms $O(e^0)$ or $O(e^1)$. They are given as  
$
S_1=-\frac{3}{4}e+O(e^3),~~
C_1=-\frac{3}{4}e+O(e^3),~~
A_1=\frac{1}{2}e+O(e^3)
$
for $n=1$-mode,
$
S_2=1-\frac{5}{2}e^2+O(e^4),~~
C_2=1-\frac{5}{2}e^2+O(e^4),~~
A_2=\frac{1}{2}e^2+O(e^4)
$
for $n=2$-mode, and 
$
S_3=-\frac{9}{4}e+O(e^3),~~
C_3=-\frac{9}{4}e+O(e^3),~~
A_3=O(e^3)
$ for $n=3$-mode.
Our basic strategy is to compare $O(e^0)$-term of $n=2$-mode
 and $O(e^1)$-term of $n=3$-mode (as we see below, the eccentricity
relevant 
for our analysis is expected to
be  small $e\lsim 0.1$). From these two 
frequencies $f_2-\delta f$ and $f_3-\delta f$ we obtain
\beq
3(f_2-\delta f)/2-(f_3-\delta f)=-\delta f/2 \label{delf},
\eeq
and thus the total mass
$m_{total}$ is estimated  for a binary with 
small eccentricity $e$ (note also
we might  determine the eccentricity  $e$ by comparing  amplitudes of two
waves). 
 Amplitude of $n=1$-mode is smaller than $n=3$-mode,  and furthermore the
 binary confusion noise would be  larger  at lower frequency \cite{lisa}.
Therefore we investigate prospect of our method by studying  
 detectability of $n=3$-mode in stead of $n=1$-mode.

\section{detectability}
The frequency difference
$\delta f$ due to periastron advance is expressed as \cite{gravitation}
\beq
\delta f=\frac{1.2\times 10^{-7}}{1-e^2}\lmk \frac{m_{total}}{2.8 M_\odot}
\rmk^{2/3}\lmk \frac{f_3}{3\times 10^{-3}{\rm Hz}} \rmk^{5/3} {\rm Hz}, \label{fpa}
\eeq
and the accumulated frequency difference
 (chirp signal) due to gravitational radiation reaction
within observational period
$T_{obs}$ is given as  \cite{peters}
\beq
({\dot f})_{GW} T_{obs}=\frac{3.0\times 10^{-8}}{(1-e^2)^{7/2}}\lmk
\frac{m_{chirp}}{1.2 
M_\odot} 
\rmk^{5/3}\lmk \frac{f_2}{2\times10^{-3}{\rm Hz}} \rmk^{11/3}
\lmk 1+\frac{73}{24}e^2+\frac{37}{96}e^4 \rmk
\lmk \frac{T_{obs}}{10{\rm yr}}\rmk   {\rm Hz}. \label{chirp}
\eeq
Thus at lower frequency the difference due to the  periastron advance 
 can take  larger value. 
The estimation error (resolution) $\Delta f$ for  wave
frequency $f$ in matched
filtering analysis is 
given by the Fisher information matrix of fitting parameters.
 For the  set of
three unknown parameters $(f, 
\dot f$, initial phase) of quasi-monochromatic wave (${\dot f}T_{obs}\ll
f$) we obtain \cite{seto} 
\footnote{When we include the direction and orientation of the source
in fitting parameters, the 
estimation errors might be somewhat larger than our evaluation. We
should also notice that these angular parameters are fixed well by
$f=2$-mode and we might not need to fit them for determining frequency
of $f=3$ mode. } 
\beq
\Delta f=4\sqrt3
\pi^{-1}T_{obs}^{-1} SNR^{-1}=6.6\times 10^{-10} \lmk
\frac{T_{obs}}{10{\rm yr}}\rmk^{-1} \lmk \frac{SNR}{10}\rmk^{-1} {\rm
Hz},  \label{fres}
\eeq
where $SNR$ is signal to noise ratio of the  wave. For the secular
frequency  
modulation $\dot f$ we have estimation error  $\Delta\dot f=3\sqrt5
\pi^{-1}T_{obs}^{-2} SNR^{-1}$.
From equations (\ref{delf})(\ref{fpa}) and (\ref{fres}) the frequency
shift $\delta 
f$ would be resolved  
\beq
\frac{\Delta f}{\delta f}\sim 0.011 (1-e^2) \lmk \frac{m_{total}}{2.8
M_\odot} 
\rmk^{-2/3}\lmk \frac{f_3}{3\times 10^{-3}{\rm Hz}} \rmk^{-5/3} \lmk
\frac{SNR}{10}\rmk^{-1}  \lmk
\frac{T_{obs}}{10{\rm yr}}\rmk^{-1} \label{pares}.
\eeq
Here (and hereafter) we consider   contribution of
 the  estimation error for $\delta f$ only  from  $n=3$-mode, as $SNR$
of $n=2$-mode 
would be larger.
In the followings we estimate the number of Galactic NBs whose
$n=3$-mode is detected with $SNR>10$.  

First we discuss time evolution of the eccentricity parameter $e$. Using
quadrupole 
formula for gravitational radiation the orbital 
 semi-major radius $a$ is related
to the eccentricity as 
$
{a}/{a_i}=(1-e_i^2)/(1-e^2) \lmk {e}/{e_i}\rmk^{12/19} 
\lkk(1+121e^2/304)/(1+121e_i^2/304)   \rkk^{870/2299} 
$
where $a_i$ and $e_i$ are their initial values \cite{peters}. At
$a/a_i\ll 1$ we 
have  an useful approximation 
$
e\sim  \lmk {a}/{a_i (1-e_i^2)} \rmk^{19/12} e_i = \lmk
{f}/{f_i} 
\rmk^{-19/18} (1-e_i^2)^{-19/12} e_i  
$
where  Kepler  law is used. For a given orbital period the
eccentricity should have
 some distribution function $F(e; P_b)$ \cite{buitrago}, but  it is poorly
known at present,  as only several NBs have detected so far \cite{brown}. 
Here we use  the binary pulsar PSR B1913+16 as a reference. This system
 has orbital 
period $P_b=2.8\times 10^4 $sec and eccentricity  $e=0.62$
\cite{1913}. From above 
approximation   its   eccentricity evolves as 
\beq
e\simeq 0.13 (f_3/0.001{\rm
Hz})^{-19/18}. \lab{eev}
\eeq
 Hereafter we use this $e$-$f_3$ relation in our order estimation.

Next we study the  frequency and spatial distribution of the Galactic
NBs.  
Assuming that the former is  in steady 
state, we can evaluate the  distribution function ${dN}/{df}$  
using the coalescence rate $R_{NS}$ of Galactic NBs as
\beq
\frac{dN}{df}=R_{NS} \lmk\frac{df}{dt}  \rmk^{-1}\sim 3.8\times 10^4 \lmk
\frac{R_{NS}}{10^{-5}{\rm yr}^{-1} } \rmk   \lmk \frac{f_2}{10^{-3}{\rm Hz}}
\rmk^{-11/3} {\rm Hz^{-1}}.
\eeq
Kalogera et al. \cite{Kalogera:2000dz} recently estimated the event rate
as $R_{NS}\simeq 
10^{-6}-5\times 10^{-4} {\rm yr}^{-1}$ (see also \cite{Phinney:1991kd}).
For spatial distribution of Galactic NBs  we
use the standard exponential disk model 
$
\rho(R,z)=\rho_0 \exp\lmk -{R}/{R_0}\rmk \exp\lmk
-{|z|}/{z_0}\rmk, 
$
where $(R,z)$ is the Galactic cylindrical coordinate
\cite{Kalogera:2000dz,hils}.  We 
fix the radial 
scale 
length $R_0=3.5$kpc and the disk scale height $z_0=500$pc,  and  assume
that the solar system exists at the position 
$R=8.5$kpc and 
$|z|=30$pc.

We have a  relation 
$h_3(f_3)\simeq- 9eh_2(2f_3/3)/4$ between $n=2$-mode and $n=3$-mode (see
eqs.[\ref{hc}][\ref{hp}]).
The  amplitude  $h_3$ of $n=3$-mode is given as
\beq
h_3\simeq4.4\times 10^{-21}\lmk \frac{m_{chirp}}{1.2
M_\odot} 
\rmk^{5/3}\lmk \frac{f_3}{3\times 10^{-3}{\rm Hz}} \rmk^{2/3}
\lmk\frac{100{\rm pc}}{r}\rmk \lmk \frac{e}{0.1}\rmk \label{amp},
\eeq
where we have taken angular average with respect to  orientation of
sources \cite{thorne}. 
From equations (\ref{eev}) and (\ref{amp}) we can estimate the distance
to a
 binary whose $n=3$-mode is
detected  with a given $SNR$, 
\beq
r=1800 \lmk \frac{f_3}{3\times 10^{-3}{\rm Hz}} \rmk^{-7/18}
 \lmk \frac{h_{rms}(f_3,T_{obs})}{10^{-23}}\rmk^{-1} \lmk \frac{10}{SNR}
\rmk  {\rm pc} ,
\eeq 
where $h_{rms}(f_3,T_{obs})$ is the noise spectrum. We use the angular
averaged sensitivity (effectively a factor of $\sqrt 5$ degradation
\cite{lisa,thorne}) to 
take  rotation of LISA into account, and adopt the noise spectrum given in
 Ref.\cite{hils00}.
When the effective frequency bin 
$T_{obs}^{-1}$ is 
occupied by 
more than one Galactic  CWDBs (more precisely, number of fitting
parameters is larger than that of  data),  their
confusion noise  becomes  important and  the
sensitivity $h_{rms}(f,T_{obs})$ would  be significantly worse at
lower frequency.  Then   binaries very close to the solar system
would be only resolved. 
This transition occurs
 at frequency   $ f_t\simeq 1.6\times 10^{-3} \lmk
{T}/{10{\rm yr}}  \rmk^{-3/11}$Hz for  abundance of 
Galactic CWDBs estimated in Ref.\cite{hils00}.
As we see below,  the prospect of 
our method depends strongly on the transition frequency and  the
position of 
$f_t$ is very important.  In an extreme model that the Galactic halo
MACHOs are  WDs, the frequency $f_t$ can be close to $\sim
10^{-2}$Hz \cite{Hiscock:2000jn} (see also \cite{Ioka:1999gf}). If  
this is true, our method would be severely limited. Spatial filters that
depends on the angular position of sources might work effectively in
some cases \cite{Hellings:2001hy}. 
 At   $f\gsim f_t$ the noise
spectrum $h_{rms}(f,T_{obs})$
is mainly   determined by the detector's intrinsic
 noise and  the confusion noise by 
extra-Galactic CWDBs  that depends on cosmological evolution of binary
systems \cite{lisa,hils,Schneider:2000sg}.
 At this frequency region  the total
 noise spectrum behaves
simply  as 
 $h_{rms}(f,T_{obs})\propto T_{obs}^{-1/2}$. In the present analysis
we only count NBs with frequency  $f_2=2f_3/3>f_t$.

Now we can estimate the number of NBs whose $n=3$-mode is detectable
by LISA with $SNR>10$. In
figure 1 we show the results for observational
period 
$T_{obs}=1$yr and $T_{obs}=10$yr. The total number (integrated in
frequency space) becomes 
$0.11(R_{NS}/10^{-5}{\rm yr})$  for 1yr observation and
$3.7(R_{NS}/10^{-5}{\rm yr})$ for 10yr. Thus our method would be
effective for a  long (but realistic) observational period
\cite{lisa}. For these NBs ($T_{obs}= 10$yr, $SNR\le
10$ and $f_2>f_t$) the total
masses  $m_{total}$ would be 
resolved within a few percent  accuracy (see
eq.[\ref{pares}]).  In figure 1 we also plot all the
Galactic NBs (dashed-line). Note that our method works well for
   Galactic NBs with $f>f_t$ in the case of  $T_{obs}=10$yr.
The estimation error for the overall amplitude of wave
becomes
$\sim 0.15 (SNR/10)^{-1}$ in the LlSA band (see table
4.5 of \cite{lisa}). Thus the  eccentricity parameter $e$ 
for a  detected NB (with $SNR>10$
for $n=3$-mode) would be
measured within $\sim 15\%$ accuracy.
It has been observationally clarified that the mass of a neutron star
is $\sim 1.4M_\odot$  \cite{1999ApJ...512..288T}  (that of an
ordinal  white dwarf is $\lsim 1.2 
M_\odot$). 
If  most of  an eccentric
binaries with $m_{total}\lsim 2.8
M_\odot$ and $f\gsim 10^{-3}$Hz  are
either NS+NS or NS-WD binary \cite{hils}, we can obtain further
information of such binaries detected by our method in the  following
manner.   One of the two  stars  would be a neutron star
with  mass  $\sim 1.4 M_\odot$, and we can estimate the
 chirp mass of the system.  
Then the distance $r$ to
the binary is obtained as  proposed by Schutz 
\cite{schutz}.

Using a similar method based on the expression 
$\Delta {\dot f}$ given just after  equation (\ref{fres}), 
we have also estimated the number of Galactic  NBs whose chirp
signal ${\dot f}_{GW}$ for $n=2$-mode due to gravitational radiation
are measured within $5\%$
accuracy ($5\times 3/5\sim 3\%$
for the chirp mass eq.[\ref{chirp}]) \cite{seto}. As shown in figure
1,  all Galactic NBs with $f_2 \gsim 4\times 10^{-3}$Hz
satisfies this observational criteria for
$T_{obs}=10$yr.
There might  be $\sim 1 (R_{NS}/10^{-5}{\rm
yr^{-1}})$ NBs whose individual masses  can be determined
accurately from  observed
$m_{total}$ and $m_{chirp}$.

One might wonder whether two gravitational waves from different binaries
(mainly CWDBs) are confused as $n=2$- and $n=3$-modes of a same
eccentric source.
 Frequency distribution of 
Galactic  close white dwarf binaries (CWDBs) is estimated as
$dN/df\sim 1.4\times 10^9 (f/10^{-3}{\rm Hz})^{-11/3}{\rm Hz^{-1}}$
\cite{hils00}.  
We can specify the 
direction of a monochromatic source  with estimation error 
$\lsim 10^{-2}$sr for frequency $f\gsim 2\times 10^{-3}$Hz (in the case of
signal to noise ratio:10, table 4.5 of Ref.\cite{lisa})  
 using annual modulation of gravitational wave  due to motion of LISA
\cite{lisa,Peterseim:1997ic}. The 
mean frequency interval 
$(\delta f)_{int}$ for binaries within a  box of $\sim 10^{-2}$sr  is
given as 
$(\delta f)_{int}\sim (10^{-2}/4\pi)^{-1}(dN/df)^{-1}\sim6\times 10^{-5}
(f/4\times10^{-3}{\rm Hz})^{11/3}$Hz.  The typical frequency
difference 
due to  periastron advance $\delta f$ (eq.[\ref{fpa}]) is much
smaller than this interval and 
coincidence of  miss-identification would be $(\delta f)/(\delta
f)_{int}\sim 2\times10^{-3}(f_3/3\times10^{-3}{\rm Hz})^{-2}$. The total
number of Galactic
CWDBs within frequency $2\times 10^{-3}{\rm Hz}<f_2<4\times 10^{-3}{\rm
Hz}$  is estimated as $\sim 4\times 10^{4}$. Thus number of
miss-identified CWDB pairs would be at most  $\sim 80$. 
The estimation error of LISA for the direction of the orbital angular
momentum 
is $\sim 10^{-1}$sr ($SNR=10$, table 4.5 of Ref.\cite{lisa}). Thus there would be
only $\sim 80\times 10^{-1}/4\pi\sim 1$ miss-identified CWDBs.
It would be also possible to
distinguish  
NBs using the estimated total mass itself, as most CWDBs have total masses
smaller than 
$2.4 M_\odot$ \cite{hils}. 

Let us summarize this Letter.  We have proposed a new method  to
determine the  total masses of low 
frequency eccentric binaries form their gravitational waves. We use
effects of  
periastron advance 
on  the  frequency space of gravitational wave.  Our method would be
effective for Galactic NBs ($f\gsim 2\times 10^{-3}$Hz) with a long term
($\sim 10$yr) 
operation of LISA.

\acknowledgements
N.S. would like to thank Takashi Nakamura for enlightening suggestions.
He also thanks Kunihito Ioka, Misao Sasaki, and  Hideyuki Tagoshi for
helpful comments. 
This work was supported in part by
Grant-in-Aid of Scientific Research of the Ministry of Education,
Culture, Sports, Science and Technology  No. 0001416.


\begin{figure}[h]
 \begin{center}
 \epsfxsize=12.cm
 \begin{minipage}{\epsfxsize} \epsffile{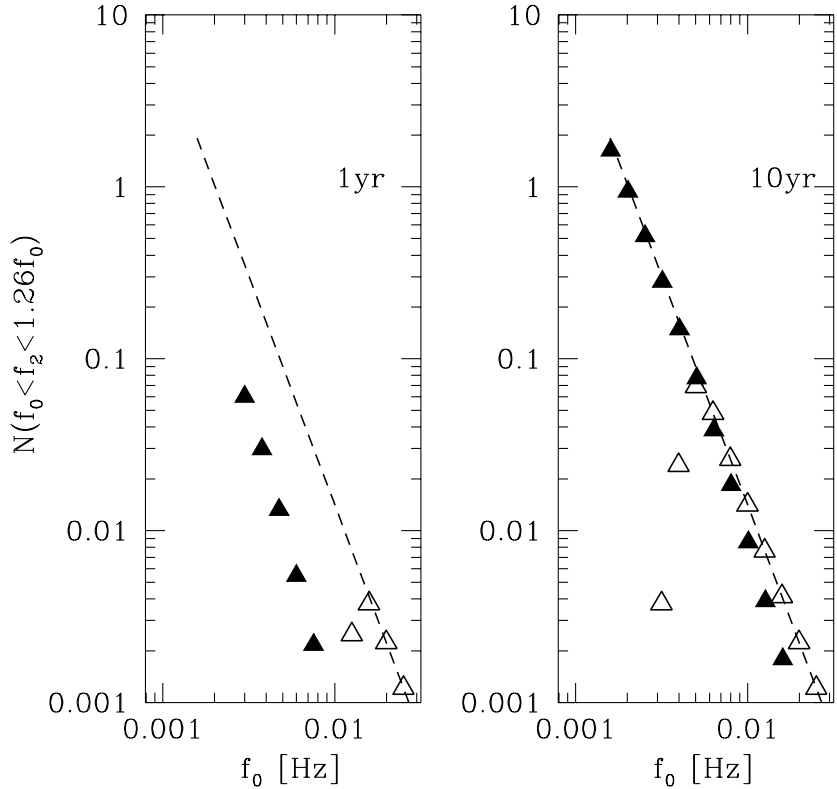} \end{minipage}
 \end{center}
\caption[]{ Various distributions of Galactic NBs within frequency bins
 $f_0\le f_2< 1.26 f_0$. The dashed-line
 represents all  Galactic NBs. The filled triangles are  the
 number 
 of NBs whose gravitational wave of
 $n=3$-mode are detected with $SNR\ge 10$, and open
 triangles are those whose chirp signals $({\dot f})_{GW}$ due to
 radiation 
 reaction  are measured  within  $5\% $ accuracy. We take the
 coalescence rate of Galactic NBs at $R_{NS}=10^{-5}{\rm yr^{-1}}$.}
\end{figure}

\end{document}